\documentclass[12pt]{iopart}

\usepackage{lineno}
\usepackage{xcolor}
\usepackage{graphicx}
\usepackage{iopams}
\usepackage{hyperref}
\hypersetup{
    colorlinks=true,   
    linkcolor=blue,    
    citecolor=blue,    
    filecolor=cyan,    
    urlcolor=blue      
}
\begin{document}
\title[The LIV effect using beyond the $\Lambda$CDM models]{Investigating the Lorentz Invariance Violation effect using different cosmological backgrounds}
\author{H. Abdalla$^{1,2,3}$, G. Cotter$^4$, M. Backes$^{5,2}$, E. Kasai$^5$, and M.~Böttcher$^2$}

\address{$^1$ IPARCOS and Department of EMFTEL, Universidad Complutense de Madrid, E-28040 Madrid, Spain}
\address{$^2$ Centre for Space Research, North-West University, Potchefstroom 2520, South Africa}
\address{$^3$ Department of Astronomy and Meteorology, Omdurman Islamic University, Omdurman 382, Sudan}
\address{$^4$ Department of Physics, University of Oxford, Keble Rd, Oxford, OX1 3RH, UK}
\address{$^5$ Department of Physics, Chemistry \& Material Science, University of Namibia, P/Bag 13301, Windhoek, Namibia}
\ead{\href{mailto:hassanahh@gmail.com}{hassanahh@gmail.com}}
\vspace{12pt}

\begin{abstract}
Familiar concepts in physics, such as Lorentz symmetry, are expected to be broken at energies approaching the Planck energy scale as predicted by several quantum-gravity theories. However, such very large energies are unreachable by current experiments on Earth. Current and future Cherenkov telescope facilities may have the capability to measure the accumulated deformation from Lorentz symmetry for photons traveling over large distances via energy-dependent time delays. One of the best natural laboratories to test Lorentz Invariance Violation~(LIV) signatures are Gamma-ray bursts~(GRBs). The calculation of time delays due to the LIV effect depends on the cosmic expansion history. In most of the previous works calculating time lags due to the LIV effect, the standard $\Lambda$CDM (or concordance) cosmological model is assumed. In this paper, we investigate whether the LIV signature is significantly different when assuming alternatives to the $\Lambda$CDM cosmological model. Specifically, we consider cosmological models with a non-trivial dark-energy equation of state ($w \neq -1$), such as the standard Chevallier-Polarski-Linder~(CPL) parameterization, the quadratic parameterization of the dark-energy equation of state, and the Pade parameterizations. We find that the relative difference in the predicted time lags is small, of the order of at most a few percent, and thus likely smaller than the systematic errors of possible measurements currently or in the near future.

\vspace{2pc}
\end{abstract}
\submitto{\CQG}
%
%
%
\vspace{-3mm}
\section{Introduction} \label{sec:intro}
\vspace{-3mm}
Over the last few decades, astronomical observations and laboratory experiments have revealed several results in fundamental physics, astrophysics, and cosmology which are in conflict with our traditional view of fundamental concepts in physics at different scales, from very small, sub-nuclear scales to very large cosmological scales~\cite{Riess98, Furniss13, Archambault14, Korochkin2019, Abdalla19h, Troitsky21}. { This has led to different theories in quantum gravity as well as various cosmological models beyond $\Lambda$CDM}~\cite{Capozziello13, wanasde14, Dominguez11b, Dzhatdoev17, W14, Nashed14, Abebe2020, bak21, Davies21, Rose, Salucci2020, Duniya2022}. {At energies close to the Planck energy scale, $E_{\rm P} \sim 1.2 \times 10^{19}$~GeV, some of the fundamental concepts in physics such as Lorentz symmetry can be broken as predicted by various quantum-gravity theories~\cite{Abdalla19h, Amelino, Ganguly2014, Tavecchio16, Lorentz17, Astapov2019, Dzhatdoev19, LI21, Terzi2021, Wei2021, Zhou:2021ycu, Wei:2021vvn, Desai:2023rkd, Sarker2023, Zheng2023, Sarker:2023mlz}. Such very large energies are unreachable by currently available facilities and instrumentation. However, current and future Cherenkov telescope facilities may have the capability to measure the accumulated deformation from Lorentz symmetry for photons traveling over large distances (cosmological scales).}\\
 The LIV effect can be described by a modification of the standard form of the relativistic dispersion relations for particles and photons~\cite{Abdalla19h, Amelino, Tavecchio16, Lorentz17, Fairbairn14, Abdalla18, Lang19, Heros2022,AZZAM2003, Carmona2023}. { Due to the LIV effect, the propagation of very-high energy~(VHE) gamma-ray photons may be slower compared to their low-energy counterparts, since higher-energy gamma-rays are expected to interact more strongly with the so-called spacetime foam}~\cite{Amelino, Tavecchio16, Lorentz17, Huerta20, Abdalla21}. { Therefore, at high energy, the refractive index for photons in vacuum is expected to be energy-dependent. This effect would lead to time delays between VHE photons of different energies received from a distant astrophysical source, which could potentially be measurable~\cite{Amelino, Tavecchio16, Wei17, Lorentz17}. 
Some of the best natural laboratories to test LIV signatures are GRBs~\cite{Abdalla19h, AZZAM2003, AZZAM2009}. 
The calculation of time delays and other cumulative LIV effects naturally depends on the cosmic expansion history.}
In almost all previous works calculating the time lag due to the LIV effect, the standard $\Lambda$CDM (or concordance) model is assumed.\\
{The simple standard $\Lambda$CDM Big Bang cosmological model (also called 'concordance model') is the most widely accepted and used cosmological model because it has been successful in interpreting a wide range of observations using several cosmological and astrophysical observables, such as the Cosmic Microwave Background~(CMB) radiation, the Big Bang Nucleosynthesis (BBN --- also called primordial nucleosynthesis), the formation and evolution of large scale structure in the Universe and the fact that the Universe is expanding~\cite{Riess98, Penzias65, Duniya2019}. However, the concordance model is facing several theoretical and observational challenges, including the singularity problem, the general initial conditions of the Universe or fine-tuning problem, the flatness problem, the horizon problem, and the accelerated expansion of the Universe}~\cite{W14, Wendy17, Duniya2022xcz}.
{The observational evidence from Type~Ia Supernovae for the accelerating expansion of the Universe~\cite{Riess98} motivated the notion of some form of dark energy. The nature of this dark energy is one of the hottest topics in modern cosmology, and several scenarios have been suggested~\cite{Capozziello13, W14, Nashed14, Bruce, Elmardi2016, Tripathi17, Rezaei17, Biswas20}. The most popular model for dark energy is a hypothetical energy that exerts negative pressure. It can be represented in terms of the density~$\rho$, pressure~$p$, and the four-velocity~$u$ using the relativistic hydrodynamical stress-energy tensor $T^{\mu \nu} = (\rho + p) u^{\mu} u^{\nu} + p g^{\mu \nu}$ and is characterized by the equation of state parameter $w = p/\rho$. The value of~$w$ could either be constant, as in the case of a cosmological constant ($w = -1$), or be redshift-dependent ($w = w(z)$), as in the model of dynamic dark-energy equation of state~(EoS) parameters}~\cite{Bruce, Rezaei17}.\\
Recently, there have been several attempts to constrain the LIV scale $E_{\rm LIV}$ using gamma-ray observations of different VHE gamma-ray sources such as Mrk~501~(see, e.g.,~\cite{Abdalla19h}). {The $\Lambda$CDM cosmological model has been commonly adopted in most past studies that analyzed time lags resulting from the LIV effect. However, there have been efforts to consider different cosmological backgrounds. One of the first attempts to incorporate these alternative cosmological models into LIV studies was presented in~\cite{Biesiada2007, Biesiada2009}. In their exploration, they utilized quintessence, Chaplygin gas, and brane-world cosmologies as cosmological backgrounds. The cosmic expansion history associated with the LIV time delay is derived from observational $H(z)$ data, employing model-independent Gaussian processes as detailed in~\cite{Pan20}. More recent investigations employed model-independent constraints on LIV via chronometers to assess LIV-induced delays, without assuming a fiducial cosmology, using Bayesian statistical techniques as detailed in~\cite{Agrawal2021, DU20, Desai2022}.\\
The primary objective of this study is to explore the sensitivity of the LIV effect, as gauged by the time lag of photons, to different cosmological backgrounds. Specifically, we aim to ascertain if employing beyond-$\Lambda$CDM cosmologies, combined with the improvement of data collected by present and upcoming instruments, enhances our capability to detect the LIV effect.
}
{In this paper, instead of the commonly used $\Lambda$CDM model ($w = -1$), we consider several alternative cosmological models to calculate time lags due to the LIV effect. Specifically, we consider background cosmological models with a non-trivial dark-energy equation of state ($w \neq -1$) to see if there is a significant deviation compared to the case of the $\Lambda$CDM model.}
{In \Sref{LIV} we derive the LIV effects on gamma-ray time lags, considering the concept of energy-momentum conservation, using a deformed dispersion relation for photons, and in \Sref{wz} we discuss different cosmological models with a non-trivial equation of state $w(z)$~\cite{Bruce, Rezaei17}. 
The results are presented in \Sref{res} where we compare the time lags due to the LIV effect considering various non-standard cosmological models with those resulting in the standard $\Lambda$CDM cosmology. 
In \Sref{sac}, we discuss and summarize our results. 
We assume the following cosmological parameters for $\Lambda$CDM and $w(z)$} (see~\cite{Bruce}) throughout the paper: $H_{0} = 70$~km~s$^{-1}$~Mpc$^{-1}$, $\Omega_{\rm m} = 0.3$, $\Omega_{\Lambda} = 0.7$.
For the Chevallier-Polarski-Linder~(CPL) and Pade parameterizations we used the best-fit parameters presented in Table~2 of~\cite{Rezaei17}. 
\section{Time lag due to the LIV effect}\label{LIV}
\vspace{-3mm}
{It is well-established that the speed of light inside a refractive medium is wavelength-dependent. Thus, photons with different wavelengths propagate at different speeds inside a refractive medium. For visible light, blue (short wavelength) light propagates more slowly compared to red light (long wavelength) inside a dispersive material. In analogy to this, various quantum gravity~(QG) theories predict that {VHE} gamma-ray photons could be sensitive to quantum fluctuations of spacetime or the so-called quantum foam, which can lead to Lorentz Invariance Violation. Therefore, the propagation of VHE gamma-ray photons is expected to be slower than that of lower-energy gamma-ray photons produced at the same distant astrophysical source (e.g.,~\cite{Abdalla19h, Tavecchio16, Lorentz17}).\\ 
In this article, we assume a flat universe with zero spatial curvature~($k=0$), as derived from observations (see, e.g.,~\cite{Planck20}).  The metric for a flat Friedmann-Robertson-Walker~(FRW) universe is written as $ds^2 = - dt^2 + a(t)^2 (dx^2 + dy^2 + dz^2)$, where $a(t)$ is the scale factor of the universe. The presence of the time-dependent scale factor $a(t)$ indicates that space-time is still curved, even though the spatial curvature is zero.
Therefore, strictly speaking, one should utilize an enhanced modified dispersion relation that includes higher-order terms, taking into account the curvature of spacetime. Recent studies have explored various approaches to analyzing the impacts of modifications to the dispersion relation in FRW spacetimes~\cite{Rosati2015, Amelino-Camelia:2023}. They examined several intriguing scenarios that deviate from the well-known formula introduced by Jacob and Piran~\cite{Jacob2008b} in the context of LIV, thereby setting the stage for new phenomenological investigations.\\
Using a modified dispersion relation in curved spacetimes is expected to produce more accurate results than adopting a flat-spacetime relation, but it significantly complicates the mathematical formalism. Straightforward spacetimes, such as Minkowski space, are easier to handle, especially when introducing new and unconventional ideas. The LIV effect is relevant at extremely high energies (near the Planck scale), where local effects may be more important than the large-scale structure of the Universe. Therefore, the difference introduced by using more complicated models is expected to be small. We defer the exploration of such effects to future work. Modifications of the photon energy dispersion relation can be represented by both superluminal and subluminal scenarios, subluminal meaning decreasing photon speed $v_{\rm ph}$ with decreasing wavelength and superluminal meaning increasing photon speed with decreasing wavelength. We can write the modifications of the photon and electron dispersion relation as~\cite{Abdalla19h, Alves2021}:
\begin{equation}\label{dis}
 E^{2} = p^2 c^2 + m^2 c^4  + ~ S   E^2 \left(\frac{E}{ E_{\rm LIV}} \right)^{n},
\end{equation}
where $S = -1$ represents a subluminal scenario, and $S = +1$ represents the superluminal case. The characteristic energy $E_{\rm LIV}$ is parameterized as a fraction of the Planck energy, $E_{\rm LIV} = E_{\rm P} / \xi_n$, where the order of the leading correction~$n$ and the dimensionless parameter $\xi_n$ depend on the theoretical framework~\cite{Abdalla19h, Amelino,  Tavecchio16}. A value of $E_{\rm LIV} \sim E_{\rm P}$ (i.e., $\xi_1 = 1$)
has been considered to be the physically best motivated choice~\cite{Tavecchio16, Liberati09}, which is also consistent with the results of~\cite{BW15} who constrained  $E_{\rm LIV} >  0.65 ~ E_{\rm P}$. For photons,~\eref{dis} simplifies to
\begin{equation}\label{3}  
 E^{2} = p^2 c^2  + S  \frac{E^{3}}{E_{\rm LIV}},
\end{equation}
when considering only the leading correction, $n=1$. From~\eref{3}, the photon speed $v_{\rm ph} = dE/dp$ in the limit $E/E_{\rm LIV} \ll 1$ can be calculated as~\cite{Abdalla19h, Lorentz17}:
\begin{equation}\label{3vs}  
 v_{\rm ph}  \approx c \left(1 + S \frac{E}{E_{\rm LIV}} \right).
 \end{equation}
\Eref{3vs} indicates that the photon speed in the case of a subluminal scenario ($S =-1$) is decreasing with increasing energy and the other way around in the case of a superluminal scenario. The time lag between two photons reaching Earth depends on the redshift of the source and on cosmological parameters. Considering two photons with energies~$E_{\rm h}$ (high energy) and $E_{\rm l}$ (low energy), measured at redshift~$z = 0$,
the time lag $\Delta t_{n}$ between the two photons' arrival times can be written as (for more details see~\cite{Abdalla19h}):
\begin{equation}\label{tlag}  
\Delta t_{n} = S ~  \frac{n+1}{2} ~ \frac{E_{\rm h}^{n} - E_{\rm l}^{n}}{E_{\rm LIV}^n} \int^{z} _{0} \frac{(1 + z')^{n}}{H{(z')}} dz',
\end{equation}
where $H(z') = H_{0} \sqrt{\Omega_{\rm m} (1+z')^3 + \Omega_{\Lambda}}$ in the $\Lambda$CDM model.\\
The time lag over energy difference $\tau_{n}$ can thus be written as:
\begin{equation}\label{tlagd}  
\tau_{n} = \frac{\Delta t_{n}}{E_{\rm h} ^{n} - E_{\rm l} ^{n}}  = S ~  \frac{n+1}{2 ~ E_{\rm LIV}^n}  \int^{z} _{0} \frac{(1 + z')^{n}}{H{(z')}} dz'.
\end{equation}
{ Obviously, the time-delay due to the LIV effect depends on the cosmological model,  which is mathematically described by the redshift dependence of the Hubble parameter}~$H(z)$.\\
\section {Redshift parametrization of the equation of state parameters} \label{wz}
\vspace{-3mm}
Observations of large-scale structure, cosmic microwave background, Type~Ia supernovae, etc., all indicate that the total energy budget of the Universe is dominated by dark matter and dark energy. The $\Lambda$CDM model has proven to be the simplest model that describes dark energy to be a constant, i.e. its equation of state is independent of redshift. Although $\Lambda$CDM is a successful model in many aspects, it still has some problems. Therefore, several families of alternative parameterizations of a redshift-dependent dark energy equation of state parameter have been proposed (for more details see, for example,~\cite{Bruce, Tripathi17, Rezaei17, Biswas20}), among which different members focus on different properties of the observed Universe. It is standard to parameterize the dark energy density as $\rho_{\rm DE}(z) = H_{0}^{2} \Omega_{\rm DE}(z = 0) f(z)$~\cite{Bruce}, with
\begin{equation}\label{ar}
f(z) = \exp \left[3\int_{0}^{z} \frac{(1+w(z'))}{(1+z')} dz'\right],
\end{equation}
where $f(z)$ is the dark energy density function~\cite{Bruce, Tripathi17, Biswas20}. For the $\Lambda$CDM cosmology, $f(z) \equiv 1$. 
The Hubble parameter $H(z)$ using the redshift-dependent equation of state $w(z)$ can be written as~\cite{Bruce, Tripathi17},
\begin{equation}\label{hd}
H(z)_{\rm DE} = H_{0} \sqrt{\Omega_{\rm m}(1+z)^3 + \Omega_{\rm DE} f(z) + \Omega_{\rm rad}(1+z)^4}.
\end{equation} 
The radiation component, $\Omega_{\rm rad}(1+z)^4$ is negligibly small for moderate redshifts and can be neglected in the further discussion.
\subsection { Quadratic evolution of the equation of state parameter~$w(z)$ }
Cosmological models with a dynamic dark energy equation of state (EoS) are distinguished from the $\Lambda$CDM cosmological model by a non-trivial equation of state ($w \neq -1$), which is redshift dependent. In the model of a quadratic evolution of the equation of state parameter $w(z)$, the EoS depends on cosmological time with values of~$w(z)$ less than or equal to zero (see, e.g.,~\cite{Bruce}).
As one possible model, we consider a quadratic parameterization of the dark energy EoS~\cite{Bruce}:
\begin{equation}\label{wquad}
w(z) = w_{0} + w_{1}z + w_{2}z^2.
\end{equation}
The equation of state today is $w(z = 0) \equiv w_{0} = -1$, and at redshift~$z_{t}$, the dark energy starts to evolve like matter $w(z_t) = 0$. 
\subsection{ Pade Parameterizations}
\vspace{-3mm}
Constraints on the CPL and Pade parameterizations in comparison to the $\Lambda$CDM model have been derived in~\cite{Rezaei17}.\\
The well-known CPL parameterization of the EoS can be written as: 
\begin{equation}
w_{\rm DE} (z) =  w_{0} + w_{1} z/(1+z).
\end{equation}
The Pade~1 and Pade~2 equation of state parameterizations were developed in~\cite{Rezaei17}, 
\begin{equation}
w_{\rm DE}^I (z) =  \frac{w_{0}  + (w_{0} + w_{1}) z}{1+(1+w_{2})z}, 
\end{equation} 
and
\begin{equation}
w_{\rm DE}^{II} (z) =  \frac{w_{0}  + w_{1} \ln{a}}{1+w_{2} \ln{a}}.
\end{equation} 
where $a = 1/(1+z)$. The values of $w_{0}$, $w_{1}$ and $w_{2}$ in each case can be found in~\cite{Rezaei17}.
\section{Results}\label{res}
\vspace{-3mm}
In this section, we present the time lag resulting from the LIV when considering non-standard cosmological models with $w(z) \ne 1$, compared to the case of standard $\Lambda$CDM cosmology. 
The time lags over energy difference $\tau_{n}$ between two photons with diffident energies using the $\Lambda$CDM model are calculated using~\eref{tlagd}. 
To investigate the LIV effect with a redshift-dependent equation of state parameter $w(z)$, we modified~\eref{tlagd} using the corresponding equations from \Sref{wz}. In this work, we are not pursuing the determination of the LIV energy scale, $E_{LIV}$, through observational data. Our primary objective is to ascertain whether the time lag arising from the LIV effect is influenced by the choice of cosmological model. For our investigations, we assumed that $E_{LIV}$  is equivalent to the Planck energy, $E_P$, approximately valued at $E_{\rm LIV} = E_{\rm P}  \sim 1.2 \times 10^{19}$. Subsequently, we evaluated the time lag attributed to the LIV effect across various cosmological models, contrasting these findings with the $\Lambda$CDM model. The outcomes of this comparison are presented in \Fref{fig:ftlag}. In the top panel of \Fref{fig:ftlag} we present the time lag over energy difference as a function of redshift, while the bottom panel represents the relative deficit/surplus of the time lag $\tau_{n}$ using different forms of the equation of state $w(z)$ compared to the $\Lambda$CDM model. \Fref{fig:ftlag} illustrates that the relative deficit/surplus for the time lag using different possible equations of state $w(z)$ is very small. Even the largest surplus (in the case of Pade~2 at redshift~$z = 1.2$) is less than 4\%. \\
\begin{figure}
\centering
 \includegraphics[scale=0.865, angle=0]{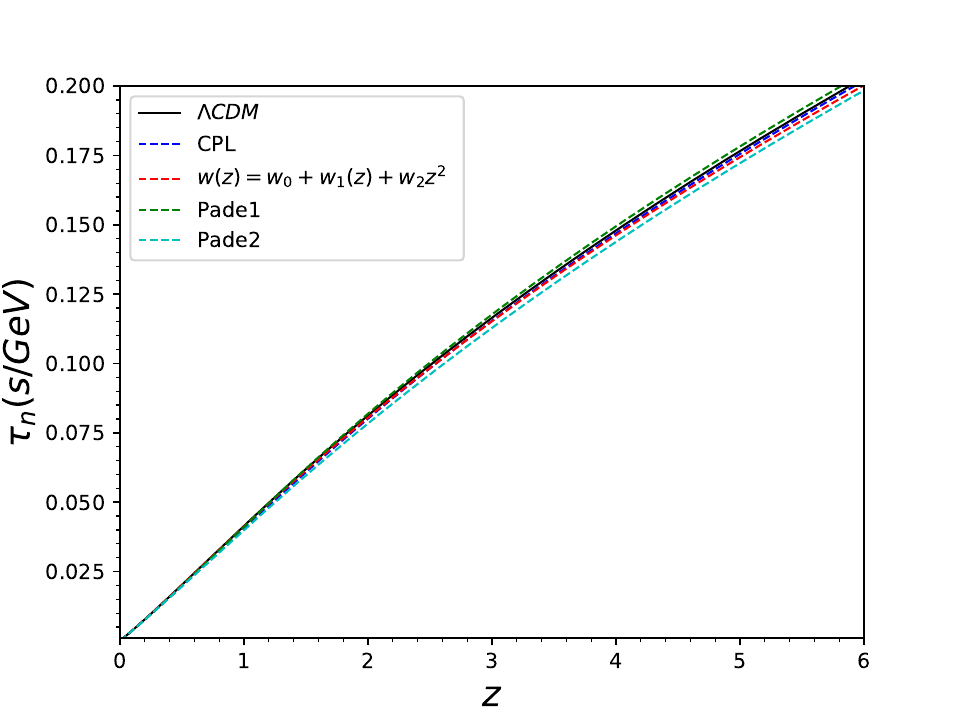}\\
 \vspace{-2mm}
 \includegraphics[scale=0.865, angle=0]{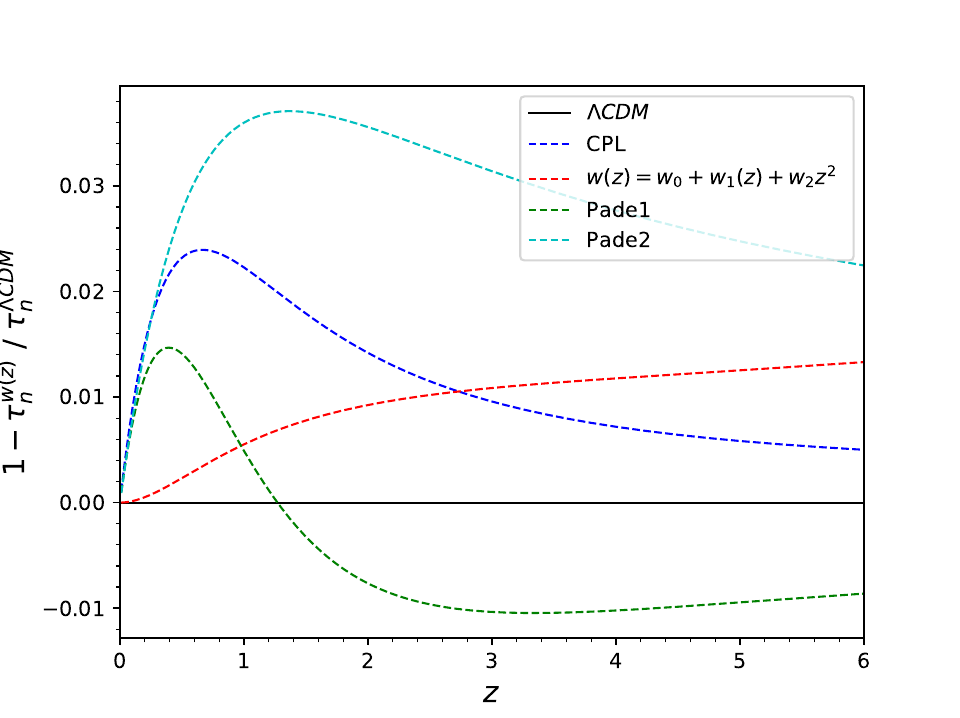}
 \caption{{\bf Top panel:} Time lag due to the LIV effect over energy difference~$\tau_{n}$ as a function of redshift using $\Lambda$CDM and models with different forms of the equation of state $w(z)$, assuming a value of $E_{\rm LIV} \sim E_{\rm P}$  .\\
 {\bf Bottom panel:} Relative deficit/surplus of the time lag using different forms of the equation of state $w(z)$ compared to the $\Lambda$CDM model.
 }
\label{fig:ftlag}
\end{figure} 
\section{Discussion and Conclusions}\label{sac}
\vspace{-3mm}
VHE gamma rays from distant astrophysical sources such as GRBs and flaring AGNs, have been used during the past few decades to constrain the LIV effect, based on the modified dispersion relation~\eref{dis}, as proposed by several quantum gravity theories. In such studies, the energy-dependent time delays as expressed by~\eref{tlag} are searched for, while usually the sources of intrinsic time delay are neglected. {In this work, our main goal was not to estimate the value of the LIV energy scale $E_{LIV}$ by comparison with observational data, but instead we investigated whether the time lag due to the LIV effect is sensitive to the underlying cosmological model. To do this, we assumed a value of $E_{\rm LIV} = E_{\rm P}  \sim 1.2 \times 10^{19}$~GeV. To determine whether the LIV effect is sensitive to the cosmological background, we computed the time lag due to the LIV effect using different cosmological backgrounds and compared the results to the $\Lambda$CDM model, as shown in \Fref{fig:ftlag}.}\\
We have presented calculations for the time lag due to the LIV effect, using non-standard cosmological models with non-trivial dark-energy equation of state parameter. The time lag between photons of different energies depends on the redshift of the source. 
The co-moving distance corresponding to a given redshift depends on the cosmological model, { as mathematically described by the redshift dependence of the Hubble parameter}~$H(z)$. 
\Fref{fig:ftlag} illustrates that the relative deficit/surplus of the time lag due to the LIV effect using different forms of the equation of state $w(z)$ compared to the $\Lambda$CDM model is very small. Several previous works have investigated similar aspects. The primary objective of~\cite{Biesiada2009} is to investigate if the LIV effect might result from an incorrect assumption of the $\Lambda$CDM cosmology in~\cite{Ellis2005}, suggesting that another model might more aptly describe the accelerating Universe. They find the effect does not diminish in alternative models. The authors determined that the effect is most pronounced in models like the quintessence model with a varying equation of state~$w(z)$, which offers the best fit for time-delay calculations. In a recent article with similar goals to the present paper~\cite{Staicova2023}, the author used two data sets (TD1 and TD2) to examine the dependence of a GRB-based time-delay estimation on the underlying cosmological model, using additional data sets of BAO and SN measurements from the Pantheon dataset. They found that the deviation between the highest and lowest results in TD1 is about~$20\%$, and can reach~$60\%$ in TD2, which indicates a very high dependence on the cosmological model. These results contradict our own findings as well as the model-independent results in~\cite{Biesiada2009, Desai2022}, who found that the time lag due to the LIV effect is insensitive to the underlying cosmological model.
We think that the reason for this discrepancy may be related to the uncertainties in the GRB models and the tension in the cosmological observations themselves. Therefore, constraining the LIV energy scale and the cosmological parameters simultaneously may not be feasibl.\\
We expect that the variations in the time lag attributed to the LIV effect across different cosmological backgrounds primarily stem from the fact that both the distance estimation and photon travel time from the source are influenced by the cosmological context. Consequently, the anticipated deviation in the LIV effect is likely to align with, or even be lower than, the percentage difference observed in distance estimations across varying cosmological backgrounds. As an illustration, the deviation in distance estimation from the $\Lambda$CDM model when considering the quadratic dark energy equation of state (see~\eref{wquad}) is approximately~$2.5\%$, see~\cite{Bruce}. As illustrated in the bottom panel of \Fref{fig:ftlag}, 
the deviation in the time lag attributed to the LIV effect is nearly~50\% less than this distance estimate deviation for the same cosmological model. Therefore, it is unlikely that the deviation from the $\Lambda$CDM model in any alternative cosmological model would surpass~$10\%$. Hence, the substantial potential deviation noted in~\cite{Staicova2023} seems improbable.}\\
As evident in the bottom panel of \Fref{fig:ftlag}, the maximum deviation from $\Lambda$CDM which we found is $<4\%$ for a source located at around $z  = 1.5$, using the Pade~2 parameterization. The predicted deficit/surplus is too small to be measured using current imaging atmospheric Cherenkov telescope facilities.
Even though the sensitivity of the CTA will be several times higher than that of the currently leading instruments H.E.S.S., VERITAS, and MAGIC (see, e.g.,~\cite{Brown18}), it will likely still be insufficient to measure the small expected difference of~$<4\%$. Therefore, we conclude that, due to systematic uncertainties, especially possible source-intrinsic time lags, the deviation from the $\Lambda$CDM cosmology in measurements of LIV-induced time lags is undetectable by the current and near-future gamma-ray experiments.

\section*{Acknowledgements}
We are grateful to the anonymous reviewers for their insightful suggestions, which have significantly enhanced the quality of our paper. We acknowledge support from the UKRI STFC Global Challenges Research Fund project ST/S002952/1. The work of HA is partially supported by the María Zambrano grant and the AfAS 2022 Seed Research Grant. The work of MBa, MBö, and EK is partially supported by the South African Department of Science and Innovation and the National Research Foundation of South Africa~(NRF) through the South African Gamma-Ray Astronomy Programme. The work of MBa is further supported in part by the NRF (Grant Number 132264). Opinions, findings, and conclusions or recommendations expressed in any publication generated by the NRF-supported research are that of the author(s), and the NRF accepts no liability whatsoever in this regard.


\section*{ORCID iDs}
Hassan Abdalla \url{https://orcid.org/0000-0002-0455-3791}\\
Garret Cotter \url{https://orcid.org/0000-0002-9975-1829}\\
Michael Backes \url{https://orcid.org/0000-0002-9326-6400}\\
Eli Kasai \url{https://orcid.org/0000-0001-9696-7221}\\
Markus Böttcher \url{https://orcid.org/0000-0002-8434-5692}

\section*{References}
\bibliographystyle{iopart-num.bst}
\bibliography{sample631}

\providecommand{\newblock}{}
\begin{thebibliography}{10}
\expandafter\ifx\csname url\endcsname\relax
  \def\url#1{{\tt #1}}\fi
\expandafter\ifx\csname urlprefix\endcsname\relax\def\urlprefix{URL }\fi
\providecommand{\eprint}[2][]{\url{#2}}

\bibitem{Riess98}
{Riess} A~G, {Filippenko} A~V, {Challis} P, {Clocchiatti} A, {Diercks} A, {Garnavich} P~M, {Gilliland} R~L, {Hogan} C~J, {Jha} S, {Kirshner} R~P, {Leibundgut} B, {Phillips} M~M, {Reiss} D, {Schmidt} B~P, {Schommer} R~A, {Smith} R~C, {Spyromilio} J, {Stubbs} C, {Suntzeff} N~B and {Tonry} J 1998 {\em The Astronomical Journal\/} {\bf 116} 1009--1038 (\textit{Preprint} \eprint{astro-ph/9805201})

\bibitem{Furniss13}
{Furniss} A, {Williams} D~A, {Danforth} C, {Fumagalli} M, {Prochaska} J~X, {Primack} J, {Urry} C~M, {Stocke} J, {Filippenko} A~V and {Neely} W 2013 {\em The Astrophysical Journal Letters\/} {\bf 768} L31 (\textit{Preprint} \eprint{1304.4859})

\bibitem{Archambault14}
Archambault S {\em et~al.\/} (VERITAS, Fermi-LAT) 2014 {\em Astrophys. J. Lett.\/} {\bf 785} L16 (\textit{Preprint} \eprint{1403.4308})

\bibitem{Korochkin2019}
{Korochkin} A, {Rubtsov} G and {Troitsky} S 2019 {\em Journal of Cosmology and Astroparticle Physics\/} {\bf 2019} 002 (\textit{Preprint} \eprint{1810.03443})

\bibitem{Abdalla19h}
Abdalla H {\em et~al.\/} (H.E.S.S.) 2019 {\em Astrophys. J.\/} {\bf 870} 93 (\textit{Preprint} \eprint{1901.05209})

\bibitem{Troitsky21}
{Troitsky} S 2021 {\em European Physical Journal C\/} {\bf 81} 264 (\textit{Preprint} \eprint{2004.08321})

\bibitem{Capozziello13}
{Capozziello} S, {Gonz{\'a}lez} P~A, {Saridakis} E~N and {V{\'a}squez} Y 2013 {\em Journal of High Energy Physics\/} {\bf 2013} 39 (\textit{Preprint} \eprint{1210.1098})

\bibitem{wanasde14}
{Wanas} M~I and {Hassan} H~A 2014 {\em BSG Proceedings 21\/}  201--208 (\textit{Preprint} \eprint{1209.6218})

\bibitem{Dominguez11b}
{Dom{\'\i}nguez} A, {S{\'a}nchez-Conde} M~A and {Prada} F 2011 {\em Journal of Cosmology and Astroparticle Physics\/} {\bf 2011} 020 (\textit{Preprint} \eprint{1106.1860})

\bibitem{Dzhatdoev17}
{Dzhatdoev} T~A, {Khalikov} E~V, {Kircheva} A~P and {Lyukshin} A~A 2017 {\em Astronomy and Astrophysics\/} {\bf 603} A59 (\textit{Preprint} \eprint{1609.01013})

\bibitem{W14}
{Wanas} M~I and {Hassan} H~A 2014 {\em International Journal of Theoretical Physics\/} {\bf 53} 3901--3909

\bibitem{Nashed14}
{Nashed} G~G~L 2014 {\em Advances in High Energy Physics\/} {\bf 2015} arXiv:1403.6937 (\textit{Preprint} \eprint{1403.6937})

\bibitem{Abebe2020}
{Abebe} A, {Al Ajmi} M, {Elmardi} M, {Nandan} H and {Ul Sabah} N 2021 {\em International Journal of Geometric Methods in Modern Physics\/} {\bf 18} 2150192-327 (\textit{Preprint} \eprint{2003.09441})

\bibitem{bak21}
{Bakry} M~A and {Shafeek} A~T 2021 {\em Gravitation and Cosmology\/} {\bf 27} 89--104

\bibitem{Davies21}
{Davies} J, {Meyer} M and {Cotter} G 2021 {\em Physical Review D\/} {\bf 103} 023008 (\textit{Preprint} \eprint{2011.08123})

\bibitem{Rose}
{Rose} S~J and {Hatfield} P~W 2021 {\em Contemporary Physics\/} {\bf 62} 14--23

\bibitem{Salucci2020}
Salucci P {\em et~al.\/} 2021 {\em Front. in Phys.\/} {\bf 8} 603190 (\textit{Preprint} \eprint{2011.09278})

\bibitem{Duniya2022}
Duniya D~G~A, Abebe A, de~la Cruz-Dombriz A and Dunsby P~K~S 2022 {\em Mon. Not. Roy. Astron. Soc.\/} {\bf 518} 6102--6113 (\textit{Preprint} \eprint{2210.09303})

\bibitem{Amelino}
{Amelino-Camelia} G, {Ellis} J, {Mavromatos} N~E, {Nanopoulos} D~V and {Sarkar} S 1998 {\em Nature\/} {\bf 393} 763--765 (\textit{Preprint} \eprint{astro-ph/9712103})

\bibitem{Ganguly2014}
{Ganguly} A~K and {Jaiswal} M~K 2014 {\em Physical Review D\/} {\bf 90} 026002 (\textit{Preprint} \eprint{1608.07994})

\bibitem{Tavecchio16}
{Tavecchio} F and {Bonnoli} G 2016 {\em Astronomy and Astrophysics\/} {\bf 585} A25 (\textit{Preprint} \eprint{1510.00980})

\bibitem{Lorentz17}
{Lorentz} M and {Brun} P 2017 {Limits on Lorentz invariance violation at the Planck energy scale from H.E.S.S. spectral analysis of the blazar Mrk 501} {\em European Physical Journal Web of Conferences\/} ({\em European Physical Journal Web of Conferences\/} vol 136) p 03018 (\textit{Preprint} \eprint{1606.08600})

\bibitem{Astapov2019}
Astapov K, Kirpichnikov D and Satunin P 2019 {\em Journal of Cosmology and Astroparticle Physics\/} {\bf 04} 054 (\textit{Preprint} \eprint{1903.08464})

\bibitem{Dzhatdoev19}
{Dzhatdoev} T, {Khalikov} E, {Podlesnyi} E and {Telegina} A 2019 {Intergalactic {\ensuremath{\gamma}}-ray propagation: basic ideas, processes, and constraints} {\em Journal of Physics Conference Series\/} ({\em Journal of Physics Conference Series\/} vol 1181) p 012049 (\textit{Preprint} \eprint{1810.06200})

\bibitem{LI21}
{Li} H and {Ma} B~Q 2021 {\em Journal of High Energy Astrophysics\/} {\bf 32} 1--5 (\textit{Preprint} \eprint{2105.06647})

\bibitem{Terzi2021}
{Terzi{\'c}} T, {Kerszberg} D and {Stri{\v{s}}kovi{\'c}} J 2021 {\em Universe\/} {\bf 7} 345 (\textit{Preprint} \eprint{2109.09072})

\bibitem{Wei2021}
{Wei} J~J and {Wu} X~F 2022 {Tests of Lorentz Invariance} {\em Handbook of X-ray and Gamma-ray Astrophysics\/} ed C B and {Santangelo} A (Springer Singapore) p~82 (\textit{Preprint} \eprint{2111.02029})

\bibitem{Zhou:2021ycu}
Zhou Q~Q, Yi S~X, Wei J~J and Wu X~F 2021 {\em Galaxies\/} {\bf 9} 44 (\textit{Preprint} \eprint{2106.05733})

\bibitem{Wei:2021vvn}
Wei J~J and Wu X~F 2021 {\em Front. Phys.\/} {\bf 16} 44300 (\textit{Preprint} \eprint{2102.03724})

\bibitem{Desai:2023rkd}
{Desai} S 2023 {\em {Astrophysical and Cosmological Searches for Lorentz Invariance Violation}\/} (Springer Singapore) submitted. (\textit{Preprint} \eprint{2303.10643})

\bibitem{Sarker2023}
{Sarker} A, {Medhi} A and {Devi} M~M 2023 {\em arXiv e-prints\/} arXiv:2302.10456 (\textit{Preprint} \eprint{2302.10456})

\bibitem{Zheng2023}
{Zheng} Y~G, {Kang} S~J, {Zhu} K~R, {Yang} C~Y and {Bai} J~M 2023 {\em Physical Review D\/} {\bf 107} 083001 (\textit{Preprint} \eprint{2211.01836})

\bibitem{Sarker:2023mlz}
Sarker A, Medhi A and Devi M~M 2023 {\em Eur. Phys. J. C\/} {\bf 83} 592 (\textit{Preprint} \eprint{2302.10456})

\bibitem{Fairbairn14}
{Fairbairn} M, {Nilsson} A, {Ellis} J, {Hinton} J and {White} R 2014 {\em Journal of Cosmology and Astroparticle Physics\/} {\bf 2014} 005 (\textit{Preprint} \eprint{1401.8178})

\bibitem{Abdalla18}
{Abdalla} H and {B{\"o}ttcher} M 2018 {\em The Astrophysical Journal\/} {\bf 865} 159 (\textit{Preprint} \eprint{1809.00477})

\bibitem{Lang19}
{Lang} R~G, {Mart{\'\i}nez-Huerta} H and {de Souza} V 2019 {\em Physical Review D\/} {\bf 99} 043015 (\textit{Preprint} \eprint{1810.13215})

\bibitem{Heros2022}
{P{\'e}rez de los Heros} C and {Terzi{\'c}} T 2022 {\em arXiv e-prints\/} arXiv:2209.06531 (\textit{Preprint} \eprint{2209.06531})

\bibitem{AZZAM2003}
Azzam W and Hasan A 2023 {\em Journal of Applied Mathematics and Physics\/} {\bf 11} 2179--2184 \urlprefix\url{https://www.scirp.org/journal/paperinformation.aspx?paperid=126849}

\bibitem{Carmona2023}
Carmona J~M, Cort\'es J~L and Reyes M~A 2023  (\textit{Preprint} \eprint{2310.12661})

\bibitem{Huerta20}
{Mart{\'\i}nez-Huerta} H, {Lang} R~G and {de Souza} V 2020 {\em Symmetry\/} {\bf 12} 1232

\bibitem{Abdalla21}
Abdalla H {\em et~al.\/} (CTA) 2021 {\em JCAP\/} {\bf 02} 048 (\textit{Preprint} \eprint{2010.01349})

\bibitem{Wei17}
{Wei} J~J, {Zhang} B~B, {Shao} L, {Wu} X~F and {M{\'e}sz{\'a}ros} P 2017 {\em The Astrophysical Journal Letters\/} {\bf 834} L13 (\textit{Preprint} \eprint{1612.09425})

\bibitem{AZZAM2009}
Azzam W, Alothman M and Guessoum N 2009 {\em Advances in Space Research\/} {\bf 44} 1354--1358 ISSN 0273-1177 \urlprefix\url{https://www.sciencedirect.com/science/article/pii/S0273117709005377}

\bibitem{Penzias65}
{Penzias} A~A and {Wilson} R~W 1965 {\em The Astrophysical Journal\/} {\bf 142} 419--421

\bibitem{Duniya2019}
Duniya D, Moloi T, Clarkson C, Larena J, Maartens R, Mongwane B and Weltman A 2020 {\em Journal of Cosmology and Astroparticle Physics\/} {\bf 01} 033 (\textit{Preprint} \eprint{1902.09919})

\bibitem{Wendy17}
{Freedman} W~L 2017 {\em Nature Astronomy\/} {\bf 1} 0169 (\textit{Preprint} \eprint{1706.02739})

\bibitem{Duniya2022xcz}
Duniya D~G~A and Kumwenda M 2023 {\em Mon. Not. Roy. Astron. Soc.\/} {\bf 522} 3308--3317 (\textit{Preprint} \eprint{2203.11159})

\bibitem{Bruce}
{Bassett} B~A, {Brownstone} M, {Cardoso} A, {Cort{\^e}s} M, {Fantaye} Y, {Hlozek} R, {Kotze} J and {Okouma} P 2008 {\em Journal of Cosmology and Astroparticle Physics\/} {\bf 2008} 007 (\textit{Preprint} \eprint{0709.0526})

\bibitem{Elmardi2016}
Elmardi M, Abebe A and Tekola A 2016 {\em Int. J. Geom. Meth. Mod. Phys.\/} {\bf 13} 1650120 (\textit{Preprint} \eprint{1603.05535})

\bibitem{Tripathi17}
{Tripathi} A, {Sangwan} A and {Jassal} H~K 2017 {\em Journal of Cosmology and Astroparticle Physics\/} {\bf 2017} 012 (\textit{Preprint} \eprint{1611.01899})

\bibitem{Rezaei17}
{Rezaei} M, {Malekjani} M, {Basilakos} S, {Mehrabi} A and {Mota} D~F 2017 {\em The Astrophysical Journal\/} {\bf 843} 65 (\textit{Preprint} \eprint{1706.02537})

\bibitem{Biswas20}
{Biswas} P, {Roy} P and {Biswas} R 2020 {\em Astrophysics and Space Science\/} {\bf 365} 117 (\textit{Preprint} \eprint{1908.00408})

\bibitem{Biesiada2007}
Biesiada M and Piorkowska A 2007 {\em Journal of Cosmology and Astroparticle Physics\/} {\bf 05} 011 (\textit{Preprint} \eprint{0712.0937})

\bibitem{Biesiada2009}
{Biesiada} M and {Pi{\'o}rkowska} A 2009 {\em Classical and Quantum Gravity\/} {\bf 26} 125007 (\textit{Preprint} \eprint{1008.2615})

\bibitem{Pan20}
{Pan} Y, {Qi} J, {Cao} S, {Liu} T, {Liu} Y, {Geng} S, {Lian} Y and {Zhu} Z~H 2020 {\em The Astrophysical Journal\/} {\bf 890} 169 (\textit{Preprint} \eprint{2001.08451})

\bibitem{Agrawal2021}
{Agrawal} R, {Singirikonda} H and {Desai} S 2021 {\em Journal of Cosmology and Astroparticle Physics\/} {\bf 2021} 029 (\textit{Preprint} \eprint{2102.11248})

\bibitem{DU20}
{Du} S~S, {Lan} L, {Wei} J~J, {Zhou} Z~M, {Gao} H, {Jiang} L~Y, {Zhang} B~B, {Liu} Z~K, {Wu} X~F, {Liang} E~W and {Zhu} Z~H 2021 {\em The Astrophysical Journal\/} {\bf 906} 8 (\textit{Preprint} \eprint{2010.16029})

\bibitem{Desai2022}
Desai S, Agrawal R and Singirikonda H 2023 {\em Eur. Phys. J. C\/} {\bf 83} 63 (\textit{Preprint} \eprint{2205.12780})

\bibitem{Planck20}
Aghanim N {\em et~al.\/} (Planck) 2020 {\em Astron. Astrophys.\/} {\bf 641} A6 [Erratum: Astron.Astrophys. 652, C4 (2021)] (\textit{Preprint} \eprint{1807.06209})

\bibitem{Rosati2015}
Rosati G, Amelino-Camelia G, Marciano A and Matassa M 2015 {\em Phys. Rev. D\/} {\bf 92} 124042 (\textit{Preprint} \eprint{1507.02056})

\bibitem{Amelino-Camelia:2023}
{Amelino-Camelia} G, {Frattulillo} D, {Gubitosi} G, {Rosati} G and {Bedic} S 2023 {\em arXiv e-prints\/} arXiv:2307.05428 (\textit{Preprint} \eprint{2307.05428})

\bibitem{Jacob2008b}
Jacob U and Piran T 2008 {\em Journal of Cosmology and Astroparticle Physics\/} {\bf 01} 031 (\textit{Preprint} \eprint{0712.2170})

\bibitem{Alves2021}
{Alves Batista} R and {Saveliev} A 2021 {\em Universe\/} {\bf 7} 223 (\textit{Preprint} \eprint{2105.12020})

\bibitem{Liberati09}
{Liberati} S and {Maccione} L 2009 {\em Annual Review of Nuclear and Particle Science\/} {\bf 59} 245--267 (\textit{Preprint} \eprint{0906.0681})

\bibitem{BW15}
{Biteau} J and {Williams} D~A 2015 {\em The Astrophysical Journal\/} {\bf 812} 60 (\textit{Preprint} \eprint{1502.04166})

\bibitem{Ellis2005}
Ellis J~R, Mavromatos N~E, Nanopoulos D~V, Sakharov A~S and Sarkisyan E~K~G 2006 {\em Astropart. Phys.\/} {\bf 25} 402--411 [Erratum: Astropart.Phys. 29, 158--159 (2008)] (\textit{Preprint} \eprint{0712.2781})

\bibitem{Staicova2023}
{Staicova} D 2023 {\em Classical and Quantum Gravity\/} {\bf Focus on Quantum Gravity Phenomenology in the Multi-Messenger Era: Challenges and Perspectives} submitted. (\textit{Preprint} \eprint{2305.06504})

\bibitem{Brown18}
{Brown} A~M 2018 {\em Astroparticle Physics\/} {\bf 97} 69--79 (\textit{Preprint} \eprint{1711.01413})

\end{thebibliography}

\end{document}